 \documentclass[final,3p,times,twocolumn]{elsarticle}
\usepackage{amssymb}
\usepackage{lineno}
\usepackage{amsmath,amsfonts}
\usepackage{graphicx}
\usepackage{subfigure}
\usepackage[perc]{overpic} %scrivo sopra le immagini
\usepackage{gensymb}
\usepackage{xcolor,varwidth}

\journal{Nuclear Physics A}

\begin{document}

\begin{frontmatter}



\title{The FLAME laser at SPARC\_LAB}

\author[a]{F.G. Bisesto}
\author[a]{M.P. Anania}
\author[a]{M. Bellaveglia}
\author[a]{E. Chiadroni}
\author[b,c]{A. Cianchi}
\author[a]{G. Costa}
\author[a]{A. Curcio}
\author[a]{D. Di Giovenale}
\author[a]{G. Di Pirro}
\author[a]{M. Ferrario}
\author[a]{F. Filippi}
\author[a]{A. Gallo}
\author[a]{A. Marocchino}
\author[a]{R. Pompili}
\author[a,d]{A. Zigler}
\author[a]{C. Vaccarezza}

\address[a]{INFN Laboratori Nazionali di Frascati, Via Enrico Fermi 40, 00044 Frascati, Italy}
\address[b]{University of Rome Tor Vergata, Via Orazio Raimondo 18, 00173 Rome, Italy}
\address[c]{INFN-Roma Tor Vergata, Via della Ricerca Scientifica 1, 00133 Roma, Italy}
\address[d]{Racah Institute of Physics, Hebrew University, 91904 Jerusalem, Israel}

\begin{abstract}
FLAME is a high power laser system installed at the SPARC\_LAB Test Facility in Frascati (Italy). The ultra-intense laser pulses are employed to study the interaction with matter for many purposes: electron acceleration through LWFA, ion and proton generation exploiting the TNSA mechanism, study of new radiation sources and development of new electron diagnostics.
In this work, an overview of the FLAME laser system will be given, together with recent experimental results.
\end{abstract}

\begin{keyword}

High power laser \sep femtosecond laser \sep plasma acceleration \sep electromagnetic source



\end{keyword}

\end{frontmatter}

%\linenumbers

\section{Introduction}
Plasma accelerating structures allow to reach high energy e$^-$ and p$^+$ beams in very short distances, thanks to the extremely high gradients achievable, up to TV/m. This makes feasibile the realization of relatively compact accelerators and also open the way to a new electromagnetic (EM) radiation sources. For this purpose, high power (TW and PW class systems), ultra-short (pulses at femtosecond scale) and ultra-intense (I$>10^{18}$ W/cm$^2$) lasers are needed to create and to use plasmas for charged particle acceleration.

In this context, the Frascati Laser for Acceleration and Multidisciplinary Experiments (FLAME) has been installed in the SPARC\_LAB Test Facility \cite{ferrario2013sparc_lab} to study the interaction with solid state matter and gases at high intensities, up to $10^{20}$ W/cm$^2$.
Indeed, in this regime, it is possible to stimulate highly non-linear plasma wakefields in order to accelerate electrons in very short distances, achieving accelerating gradients greater than $100$ GV/m. On the other hand, also proton and light ion acceleration can be realized through the interaction with solid matter.

In this paper, an overview of the high power laser FLAME will be given as well as a summary of recent experimental results.

\section{The FLAME laser}
\begin{figure}[htb!]
\centering
\includegraphics[width=0.9\columnwidth]{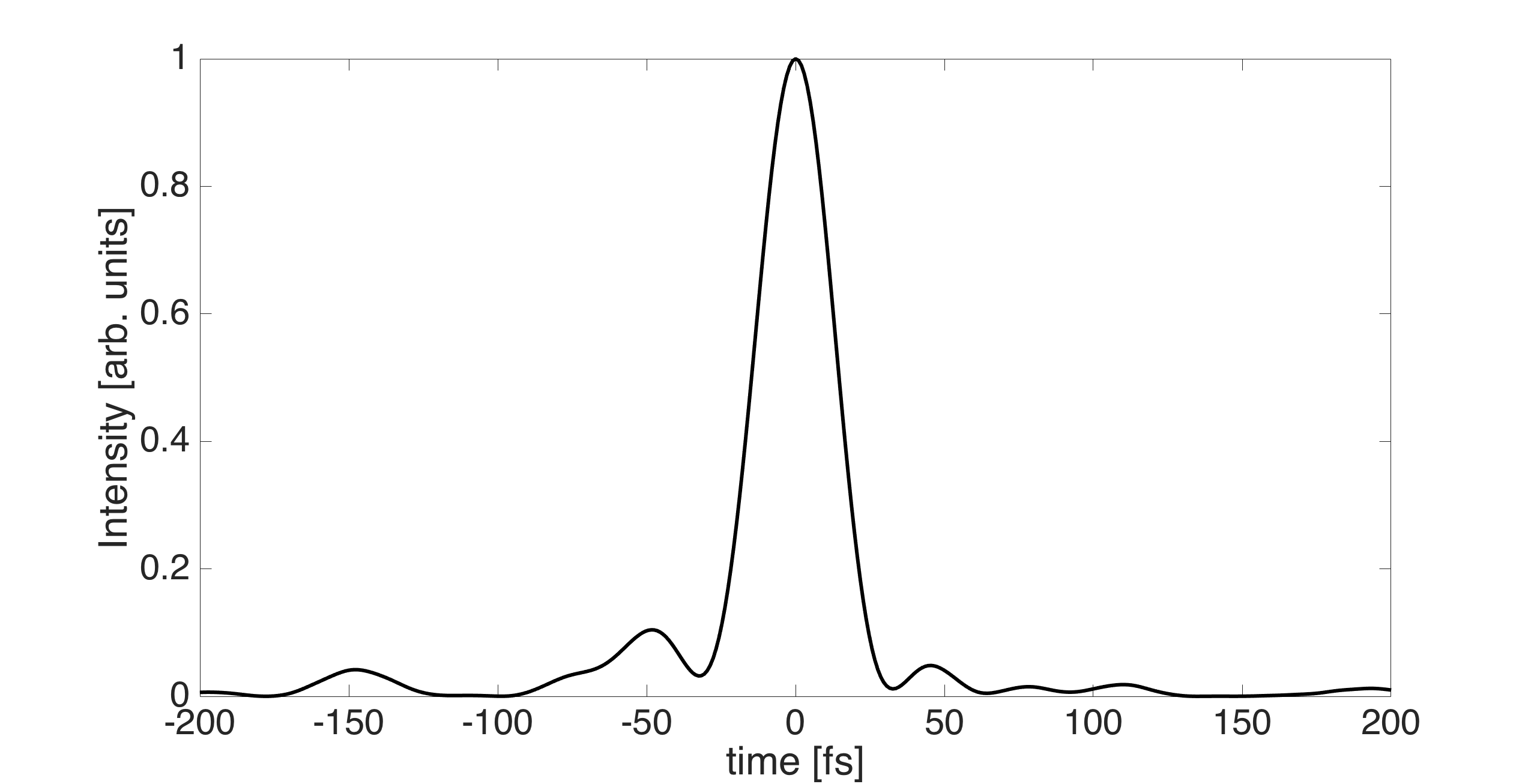}
\caption{FLAME temporal profile measured with an APE SPIDER. The temporal length is $25$ fs full width half maximum (FWHM).}
\label{fig:spider}
\end{figure}

\begin{figure*}[htb!]
\centering
\includegraphics[width=0.9\textwidth]{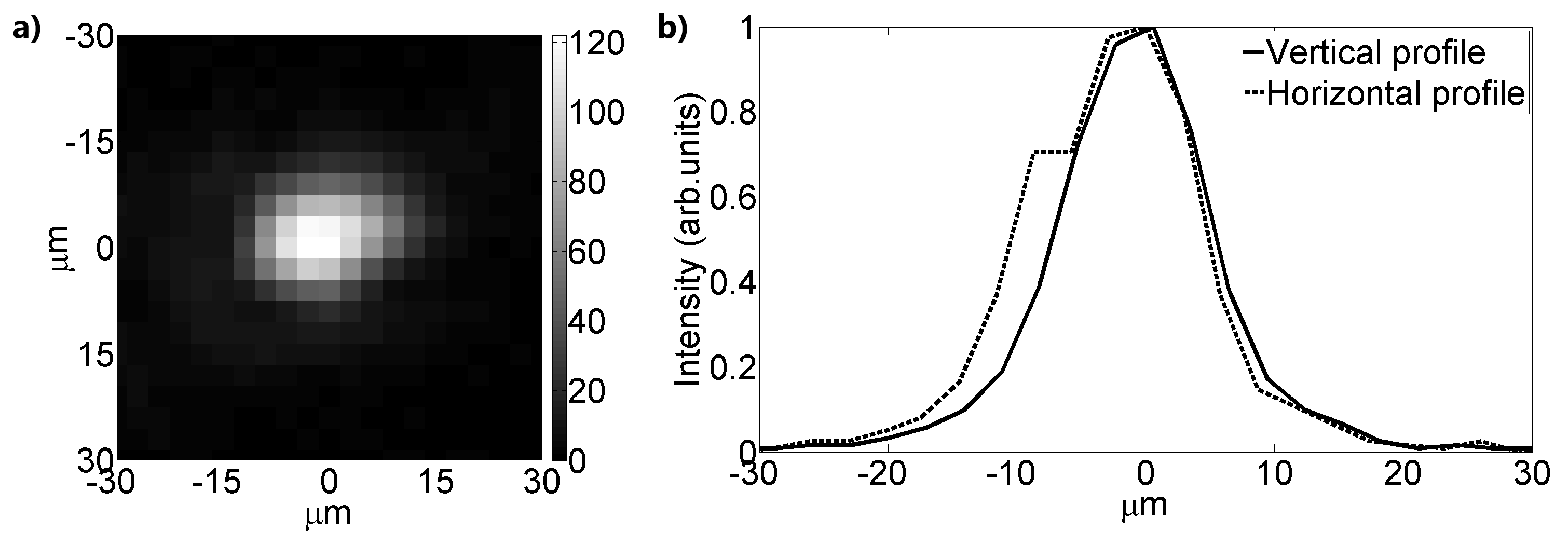}
\caption{\textbf{a)} Transverse spot size in the target area interaction point performed with a Basler Scout scA640-70gm CCD camera equipped with a $35$ mm microscope objective resulting in a resolution of $3\ \mu$m. \textbf{b)} The transverse line profile is gaussian with $\sigma=5\ \mu$m.}
\label{fig:spot}
\end{figure*}

FLAME is a compact femtosecond laser system able to provide pulses with a maximum energy of $7$ J, temporally compressed down to $25$ fs, with a peak power of $200$ TW and $10$ Hz repetition rate. The system is a Titanium-Sapphire laser based on the so-called \emph{Chirped Pulse Amplification} (CPA) scheme \cite{strickland1985compression}.  
The CPA technique involves temporal stretching of the ultra-short pulse delivered by an oscillator in order to safely amplify the pulses in solid states materials. After amplification, the laser pulse is compressed back to a value as close as possible to the initial one. In this way, it is possible to obtain a high intensity ultra short pulse.
\begin{figure*}[htb!]
\centering
\includegraphics[width=0.95\textwidth]{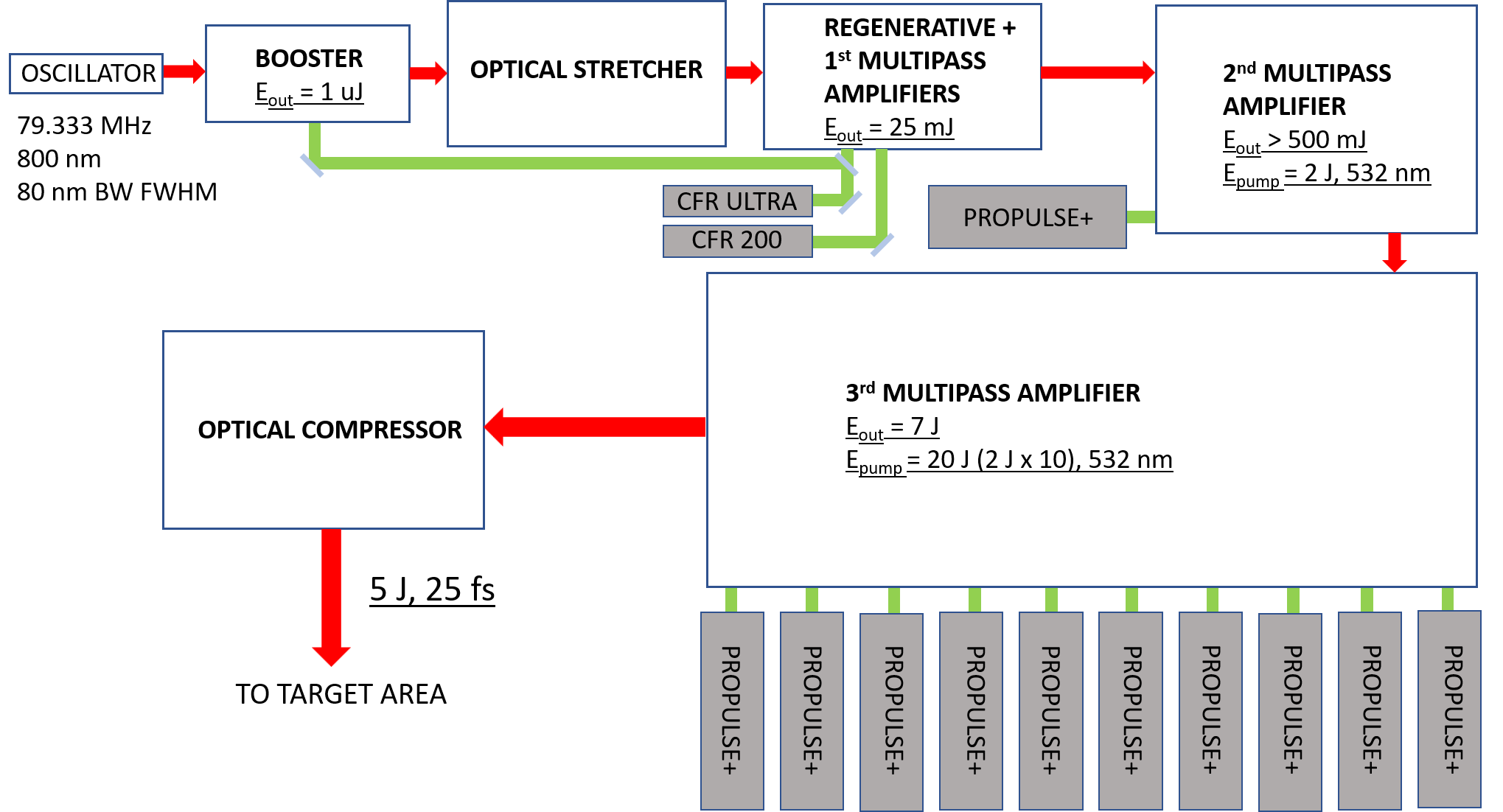}
\caption{FLAME laser optical layout.}
\label{fig:flamesetup}
\end{figure*}
The FLAME laser CPA chain (Fig.\ref{fig:flamesetup}) starts with a femtosecond laser oscillator, delivering $10$ fs, $4$ nJ pulses at $80$ MHz. A Pockels cell, acting as pulse picker, brings the oscillator output repetition rate down to $10$ Hz. In order to improve Amplified Spontaneous Emission (ASE) contrast ratio (less than $10^{-9}$), the beam is amplified at $\mu$J level by a compact multipass amplifier (Booster). After a saturable absorber, removing the residual ASE, the pulses are temporally elongated by means of an optical stretcher. This unit is based on an all-reflective triplet combination Offner, composed of two spherical concentric mirrors:
the first mirror is concave and the second is convex. This combination presents
interesting properties for use in a pulse stretcher. Indeed, it is characterised by a complete
symmetry, therefore only the symmetrical aberrations can appear (spherical aberration and
astigmatism). This combination has no on-axes coma and exhibits no chromatic
aberration (for more details \cite{cheriaux1996aberration}). At the exit, with a temporal length of about $600$ ps, an acousto-optic programmable dispersive filter is installed. It is used as a phase modulator to pre-compensate for dispersion and phase distortions introduced throughout the laser system. Then, the whole CPA chain starts: 1) a regenerative amplifier produces $\sim 1$ mJ TEM$_{00}$ pulses, with contrast ratio improved thanks to two Pockels cells and an acousto-optic programmable gain control filter compensates for gain narrowing of the spectrum; 2) three multi-pass amplifiers bring the energy up to $7$ J before compression. They are pumped by Nd-YAG lasers; the last amplifier is cryo-cooled to avoid thermal effects on the crystal ($50\times 50\times 20$) mm$^3$. An optical compressor, in a four-passes scheme, installed in a vacuum chamber at $10^{-6}$ mbar, is able to recompress the pulse down to $25$ fs without any residual spatial chirp. Figure \ref{fig:spider} shows a typical temporal measurement, performed by means of an APE SPIDER. 
A series of remotely controlled mirrors are used to transport the beam from the compressor up to the target area. Here, a dedicated experimental chamber is installed. The laser is focused by means of a $15^{\circ}$, gold-coated, Off-Axis Parabolic (OAP) mirror, reaching a transverse size of $\sigma=5\ \mu$m. Figure \ref{fig:spot} shows a typical transverse size measurement performed with a Basler Scout scA640-70gm CCD camera equipped with a $35$ mm microscope objective, with a resolution of $3\ \mu$m/pixel.
Besides the $200$ TW line, an ancillary laser is also available in target area. Indeed, a small amount of the main beam energy (about $10\%$) is split  before the cryo-cooled amplifier. Then, a dedicated compressor shortens the pulse down to $30$ fs, while a delay line synchronizes the two beams, compensating for the longer optical path of the main laser inside the last amplification stage. This second laser line is usually employed in pump-and-probe experiments: e.g. to measure the plasma density in laser wakefield acceleration (LWFA) runs, to detect THz field with EOS. In detail, for this purpose, the fine synchronization between the two beams is achieved by measuring the second harmonic generation occurring in an $\alpha$-cut beta barium borate (BBO) crystal.

\section{Recent experimental activities}
FLAME is employed as a tool to investigate the physics of the interaction between high power ultra-intense lasers and matter. In detail, two main activities are running by exploiting our $200$ TW laser: one concerns the acceleration of electrons through LWFA in a gas, the other aiming to generate proton and ion bunches from the interaction with solid targets. Besides to study the particle source, we tested new single-shot diagnostics to well characterize the electrons emitted during the interaction, both with gas and solid target.

\subsection{Electron acceleration through LWFA}
Nowadays, plasma wakefield acceleration is a promising acceleration technique for compact and cheap accelerators, needed in several fields, e.g., novel compact light sources for industrial and medical applications. Indeed, the high electric field available in plasma structures ($>$100 GV/m) allows for accelerating electrons at the GeV energy scale in a few centimeters \cite{geddes2004high}. 
At FLAME we are working in this direction thank to the so called Laser Wakefield Acceleration (LWFA) \cite{tajima1979laser} mechanism: a highly intense ($>10^{18}$ W/cm$^2$) femtosecond laser can induce waves into a plasma and, thanks to its ponderomotive force, it can inject some electrons that, in turn, get accelerated \cite{esarey2009physics}. This is called \emph{self-injection} regime \cite{faure2004laser,kalmykov2009electron}.
\begin{figure}
\centering
\includegraphics[width=1\columnwidth]{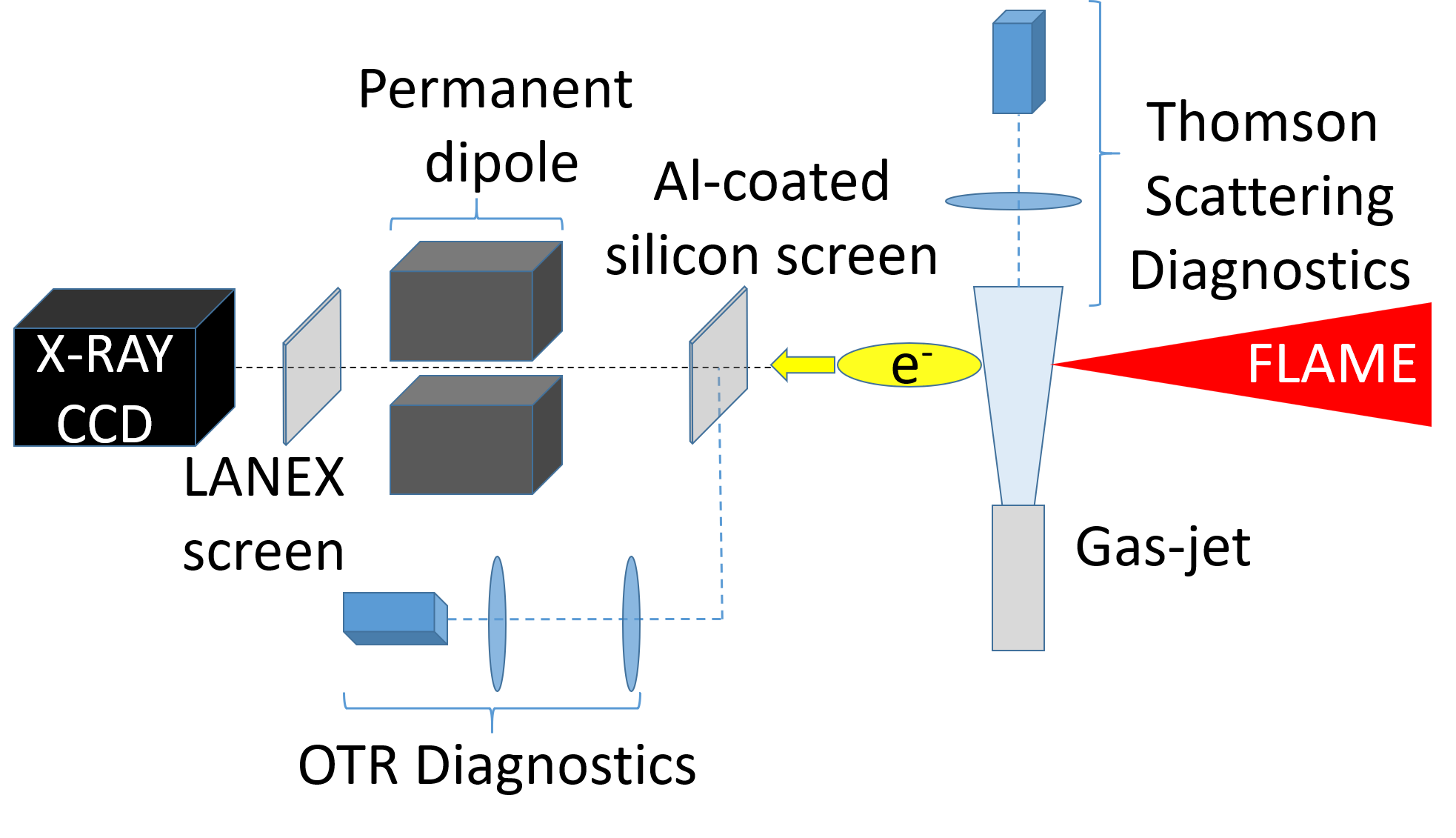}
\caption{LWFA set-up at FLAME Facility. A $2$ mm gas-jet injects N$_2$ or He$_2$ inside the interaction chamber. The laser ionizes it and plasma wakefields accelerate some electrons. The produced beam is characterized in energy, through a magnetic spectrometer, in size and divergence, by means of a LANEX screen, placed $50$ cm downstream the source. An Al-coated silicon screen is installed to measure the backward OTR produced by the electrons, while an optical system looking at the Thomson scattering along the plasma channel provides an estimation of its length. Moreover, a Mach Zehnder interferometer (not shown in this figure) allows to measure the electron density. A CCD for the X-rays produced by betatron oscillations completes the setup.}
\label{fig:selfinj}
\end{figure}
Recently, we have successfully accelerated relativistic electron bunches by employing the FLAME laser to ionize a gas (He$_2$ or N$_2$), injected in the target area by a $2$ mm super-sonic nozzle, in a highly non-linear regime. Figure \ref{fig:selfinj} shows the experimental setup realized for the LWFA experiment. 
The electron bunches have been characterized in charge, transverse size, angular divergence and energy \cite{gemma}.
In detail, we have measured typical bunches with a charge up to $10$ pC, %$\sigma=2$ mm, $\sigma^{\prime}=5$ mrad,
 $E=250$ MeV and $\Delta E=20\ \%$, while the plasma density, measured by means of a Mach-Zehnder interferometer, was equal to $n_e=1\times 10^{19}$ cm$^{-3}$. 
%Moreover, by looking at the $90^{\circ}$ Thomson scattering light, the plasma channel length was $\approx 2$ mm.

This source has been employed to study new single-shot diagnostics, exploiting both the betatron radiation emitted in the plasma channel \cite{curcio2017first} and the Transition Radiation (TR) generated by a charged particle passing through the boundary between two media with different refractive index \cite{bisesto2017novel}. 
In particular, by measuring at the same time the electron energy and the betatron X-ray spectrum, we have developed an innovative, single-shot, non-intercepting monitor of the transverse profile of plasma-accelerated electron beams \cite{curcio2017single}, allowing to perform measurement with nanometric resolution. Moreover, betatron radiation has been studied also for single-shot emittance measurements of electron bunches from LWFA inside the plasma bubble \cite{curcio2017trace}. On the other side, TR has been employed for a new single-shot emittance measurement, relying on the correlation term reconstruction by means of a microlens array \cite{bisesto2017innovative,bisesto2017innovative2,bisesto2017novel}.

Furthermore, we are working on new acceleration schemes based on plasma structures, but providing good quality electron beams. In particular, a new experimental beamline for external injection acceleration \cite{rossi2014external,rossi2016stability} is currently under development at SPARC\_LAB. In this scheme, a pre-existing electron bunch, produced by a high brightness photoinjector, is accelerated by the plasma wakefield excited by a high power laser, aiming to preserve the beam quality as much as possible. At FLAME the use of a capillary to host this interaction has been studied and a matching condition between the capillary radius and a super-Gaussian-like laser has been found \cite{bisesto2016laser}.

\subsection{Probing of fast electrons from laser-solid state matter interaction}
Ion acceleration from thin foils irradiated by high-intensity short-pulse lasers is one of the most interesting aspects in this research field since it produces a large amount of particles with energies in the multi-MeV range \cite{clark2000energetic,snavely2000intense,mackinnon2002enhancement}. 
Following theoretical models \cite{dubois2014target,wilks2001energetic,mora2003plasma}, the process starts when the electrons, directly accelerated by the laser, pass through the target. The majority of them spread out and dissipate energy, while only the fastest component can reach the target rear side \cite{singh2013direct}. Afterwards, the most energetic electrons escape, leaving an electrostatic potential on target, due to the unbalanced positive charge left on it \cite{poye2015physics}. Such a potential generates an electric field that ionizes and accelerates surface ions in a process called Target Normal Sheath Acceleration (TNSA) \cite{wilks2001energetic}.
\begin{figure}[htb!]
\centering
\includegraphics[width=0.8\columnwidth]{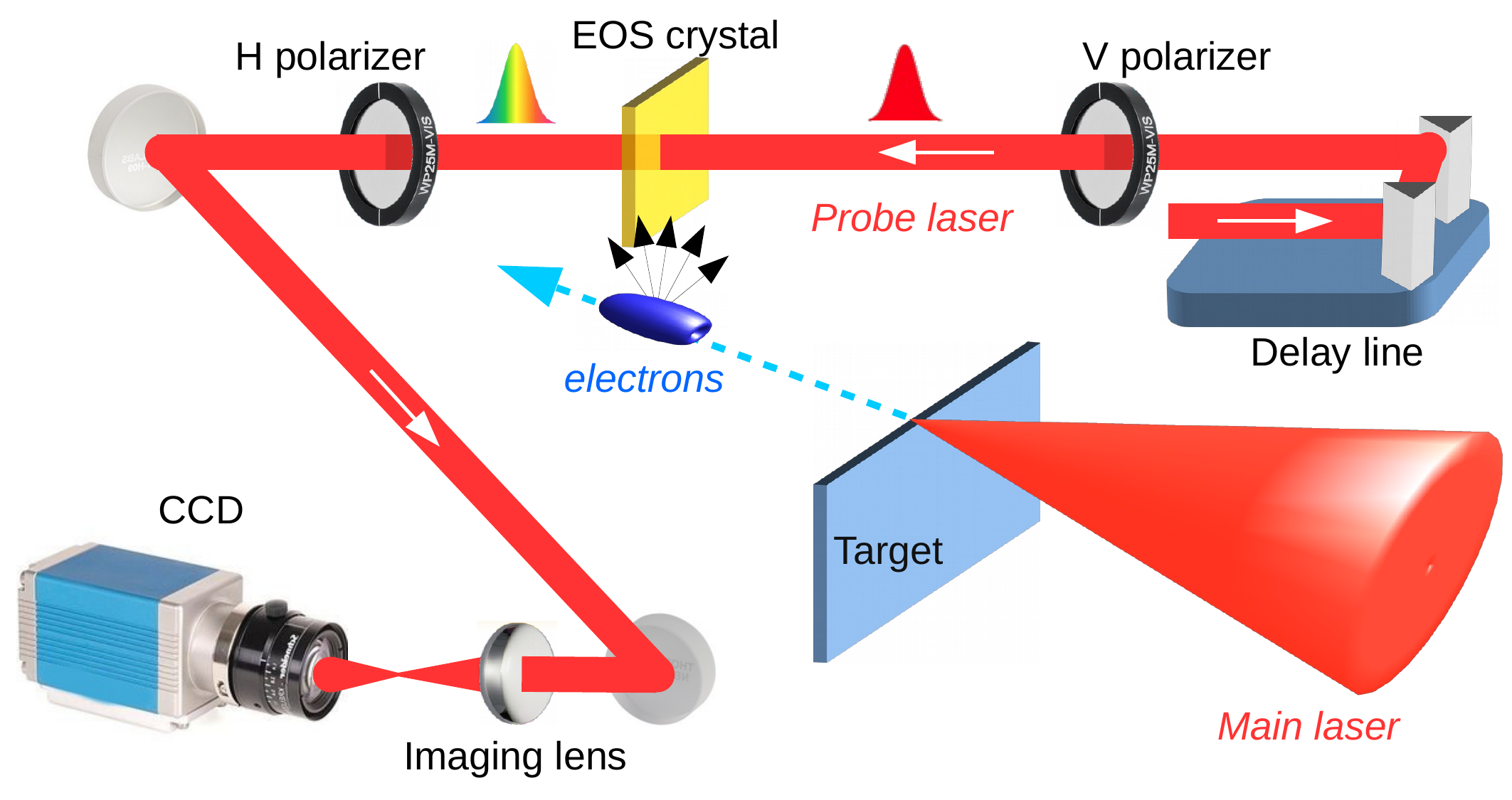}
\caption{Setup of the experiment. The FLAME laser is focused on a metallic target. The EOS diagnostics, based on a ZnTe crystal placed $1$ mm downstream the target, allows for measuring the temporal profile of the emitted electrons by means of an ancillary laser beam, directly split from the main laser, probing the local birefringence induced by the electric field.}
\label{FlameSetup}
\end{figure}
At FLAME, we have shown direct and temporally resolved measurements of the electric field carried by fast electrons.  We have used a diagnostic based on Electro-Optical Sampling (EOS) (see Fig. \ref{FlameSetup}) \cite{wilke2002single}, widely utilized in conventional accelerators \cite{steffen2009electro,eos_jitter,pompili2017electro}, with sub-picosecond resolution. Moreover, the role of the target shape in the fast electron emission has been studied, comparing planar, wedge and tip target geometries. These studies have revealed an increase in terms of charge and energy of the fast electrons, with a consequently enhancement of their electric field, using tip targets. Therefore, this result shows the possibility to boost the ion acceleration by exploiting structured targets. \cite{pompili2016femtosecond}

\section{Conclusions and perspectives}
The high power laser FLAME is an excellent tool to investigate the interaction with matter at very high intensities (up to $10^{20}$ W/cm$^2$), thanks to the efficient compression system, providing $25$ fs pulses, and an off-axis parabolic mirror to reach a tight focus ($\sigma=5\ \mu$m) in the target area. 
Thanks to the $250$ TW achievable with such a system, studies on non-linear laser wakefield acceleration have been conducted. In particular, by means of a $2$ mm gas-jet, filled with N$_2$ or He$_2$, a compact electron source has been set-up \cite{gemma} and employed to test new single-shot diagnostics \cite{curcio2017trace,bisesto2017novel}, extremely important to better understand and control plasma-based accelerators.

Moreover, FLAME is employed also to study the interaction with solid state matter. More in detail, we have designed a diagnostic based on Electro Optical Sampling (EOS) to probe the fast electron emission. In particular, its longitudinal electric field has been measured for the first time and an enhancement has been revealed by using structured targets \cite{pompili2016femtosecond}.

For the near future, a new experiment, aiming to accelerate an electron beam produced by the SPARC\_LAB high brightness photoinjector in a laser wakefield accelerator, is under design. Indeed, by guiding the FLAME laser inside a capillary filled with H$_2$ gas, it would be possible to stimulate linear wakefields accelerating the external electron bunch while preserving its initial quality, i.e. low emittance and energy spread. 
%For this purpose, a fs-level synchronization with the SPARC\_LAB photoinjector is needed.

\section*{Acknowledgments}
This work was supported by the European Union‘s Horizon 2020 research and innovation programme under grant agreement No. 653782.

\section*{References}
\bibliographystyle{elsarticle-num} 
\bibliography{eaac17_bib}


%\begin{thebibliography}{00}


%\bibitem{}
%
%\end{thebibliography}
\end{document}